# Escape of a harmonically forced particle from an infinite-range potential well: a transient resonance.

O.V.Gendelman

*Faculty of Mechanical Engineering, Technion – Israel Institute of Technology*

*Haifa 3200003 Israel*

e-mail: ovgend@technion.ac.il

The paper considers a transient process of escape of a classical particle from a one-dimensional potential well. We address a particular model of the infinite-range potential well that allows independent adjustment of the well depth and of the frequency of small oscillations. The problem can be conveniently reformulated in terms of action-angle variables. Further averaging provides a nontrivial conservation law for the slow flow. Then, one can consider the problem in terms of averaged transient dynamics on primary 1:1 resonance manifold. This simplification allows efficient analytic exploration of the escape process. As a result, one obtains a theoretical prediction for minimal forcing amplitude required for the escape, as a function of the excitation frequency. This function exhibits a single sharp minimum for a certain intermediate frequency value, below the frequency of small free oscillations. This result conforms to earlier numeric and semi-analytic estimations for similar escape models, considered, in particular, in connection with problems of ship capsize and dynamic pull-in in microelectromechanical systems. The results presented in the paper allow conjecturing the generic dynamical mechanism, responsible for these regularities. In particular, the aforementioned sharp minimum in the frequency-amplitude domain is related to formation of heteroclynic connection between the saddle points on the resonance manifold. Numeric simulations are in complete qualitative and reasonable quantitative agreement with the theoretical predictions.

1. **Introduction.**

Escape from a potential well is a classical problem, relevant in many branches of physics, chemistry and engineering [1-5]. Among many examples of possible applications, one encounters dynamics of molecules and absorbed particles, celestial mechanics and gravitational collapse, energy harvesting [6], responses of Josephson junctions [7], various aspects of capture into and escape from the resonance [8, 9], as well as a capsize of ships [3, 10]. Additional profound, and widely explored manifestation of the escape phenomenon is a dynamic pull-in in microelectromechanical systems (MEMS) [11-14].

In many of the aforementioned applications, the escape occurs due to certain type of the external forcing. One profound example of this sort is a random forcing, that can lead to widely explored phenomenon of the stochastic resonance [15, 16]. Other approach treats the constant forcing, and considers the escape due to possible slow variation of the system parameters and subsequent bifurcations of the steady-state response regimes [2].

Paper [4] adopts different setting and considers escape of a particle with zero initial conditions, after the harmonic forcing is switched on. Semi-analytical treatment includes approximate computation of the steady-state solution, based on a primary harmonic balance. Then the escape is assumed, when the instantaneous energy of the steady-state solution achieves the potential barrier. This "energy criterion" in some cases is adjusted by application of empirical correction factor. The paper considers three one-dimensional examples - potentials with negative cubic and quartic terms, and a single-mode model of buckling shallow arch. In all these examples, both direct numerics and semi-analytic energy criterion described above revealed a peculiar common feature – a sharp minimum of the critical force amplitude, necessary for the escape, at certain frequency below the frequency of small oscillations in the well. Similar energy-based escape criterion was applied for the bistable magnetic pendulum [6]. The escape curves with the sharp minimum in frequency – voltage domain were observed in the problem of dynamic pull-in in MEMS [12,13].

The energy escape criterion in [4] and some of the subsequent works, as well as the theoretical explorations of the dynamic pull-in in MEMS [11], rely on steady-state responses of the forced particle in the well in conditions of the primary resonance. This approach allows evaluation of important safety boundaries in the parametric space, but does not provide detailed understanding of the escape dynamics. The reason is clear - the escape from the well is a *transient* process by its very nature, and should be treated accordingly.

The paper is devoted exactly to this issue, and considers the escape from the potential well under the external harmonic forcing as the *transient resonance response*. We are going to demonstrate that such approach offers quantitative explanation for the regularities revealed in [4] and subsequent works. To be specific, we consider the particle trapped in a one –dimensional potential well with finite depth and rapidly decaying tails at $\pm\infty$. Similarly to [4], the particle is initially at rest at the well bottom, and at $t=0$ the external harmonic forcing with certain frequency and amplitude is switched on. The question is whether this particle will escape the trap. Formally, such escape would correspond to an infinite remotion of the particle as $t \rightarrow \infty$.

It will be demonstrated below that the process of escape in the considered model can be efficiently described by exploring the dynamics of the system on the reduced resonance manifold (RM). Critical forcing amplitude that corresponds to the escape threshold, is related to a modification of topological structure of special phase trajectories on the RM. Similar approaches received a lot of attention recently in connection with the problems of localization and energy transport in coupled oscillators [17, 18] an targeted energy transfer [19, 20]. However, the escape problem has an important peculiarity – the frequency of free oscillations of the particle in the trap varies from unity (in non-dimensional units) at the well bottom to zero at the separatrix. Therefore the approach of complexification – averaging [17, 18] applied in the aforementioned problems, and mathematically equivalent to a harmonic balance with slowly varying amplitudes [21] is hardly applicable for this transient problem. It is demonstrated below that more general formalism based on a canonlical transformation to the action-angle (AA) variables, and subsequent averaging, helps to overcome this difficulty [22]. We use the potential well in the form of squared hyperbolic secant, which allows transition to the AA variables in terms of elementary functions. This model was considered as an example of the oscillator with softening nonlinearity, exactly integrable in terms of elementary functions [23]. Among various applications of this model, it is possible to mention a problem of control over random vibrations [24]. Similar elementary integrable model of the oscillator with hardening nonlinearity, also considered in [23], was recently used for modeling the transitions from strongly to weakly nonlinear dynamics in coupled oscillators, evolution of vibration modes, and nonlinear beat phenomena [25, 26].

Section 2 of the paper is devoted to a description of the model and theoretical computation of the RM, based on the conservation law for the averaged system. In Section 3 the escape from the potential well is related to the bifurcations of the special RM phase trajectories. Section 4 comprises numeric validations and illustrations of the analytic findings. Section 5 is devoted to the concluding remarks.

2. **Description of the model and computation of the resonance manifold.**

General approach to computation of the RM in terms of the AA variables for the particle under external periodic forcing has been presented in [22]. Main details of this approach are presented below in slightly simplified form, for the sake of completeness. Let us consider the particle in the potential well under action of external harmonic forcing and without dissipation. This system can be described by the following equation:

$$\ddot{q} + \frac{\partial V}{\partial q} = F\sin\Omega\tau. \qquad (1)$$

Here $q(t)$ is a generalized dimensionless displacement of the particle, $V(q)$ is the external potential, $F$ and $\Omega$ are dimensionless amplitude and frequency of the external forcing respectively, $\tau$ is a dimensionless time. The dot denotes differentiation with respect to $\tau$. Mass of the particle is set to unity without loss of generality. System (1) can be derived from the following Hamiltonian:

$$H = H_0(p,q) - Fq\sin\Omega\tau; \; H_0 = \frac{p^2}{2} + V(q), \; p = \dot{q}. \qquad (2)$$

$H_0(p,q)$ is the component of the Hamiltonian that describes free motion of the particle in the potential well. This component can induce a transformation to the AA variables in accordance with well-known formulas [1]:

$$I(E) = \frac{1}{2\pi}\oint p(q,E)dq; \; \theta = \frac{\partial}{\partial I}\int_0^q p(q,I)dq. \qquad (3)$$

Here $H_0(p,q) = E = \text{const}$ defines a constant energy level. By inverting expressions (3), one can obtain explicit formulas for the canonical change of variables $p = p(I,\theta)$, $q = q(I,\theta)$; the conservative component of the Hamiltonian is reduced to the form $H_0 = H_0(I) = E(I)$. The canonical transformation outlined above does not include the explicit time dependence; therefore the Hamiltonian of System (1) is written in the following form in terms of the AA variables:

$$H = H_0(I) - Fq(I,\theta)\sin\Omega\tau. \tag{4}$$

Due to $2\pi$-periodicity of the angle variable, the Hamiltonian (4) can be rewritten in terms of Fourier series [27]:

$$H = H_0(I) + \frac{iF}{2}\sum_{m=-\infty}^{\infty} q_m(I)\left[\exp i(m\theta + \Omega\tau) - \exp i(m\theta - \Omega\tau)\right]; \quad q_m = q_{-m}^*. \tag{5}$$

Then, Hamilton equations will take the form:

$$\begin{aligned}
\dot{I} &= -\frac{\partial H}{\partial \theta} = \frac{F}{2}\sum_{m=-\infty}^{\infty} m q_m(I)\left[\exp i(m\theta + \Omega\tau) - \exp i(m\theta - \Omega\tau)\right] \\
\dot{\theta} &= \frac{\partial H}{\partial I} = \frac{\partial H_0}{\partial I} + \frac{iF}{2}\sum_{m=-\infty}^{\infty} \frac{\partial q_m(I)}{\partial I}\left[\exp i(m\theta + \Omega\tau) - \exp i(m\theta - \Omega\tau)\right]
\end{aligned}. \tag{6}$$

We consider the primary 1:1 resonance. To treat this regime, one *assumes* slow evolution of the phase variable $\vartheta = \theta - \Omega t$; all other phase combinations in Equations (6) should be considered as the fast phase variables. Averaging over these fast phase variables yields the following system of the slow-flow equations:

$$\begin{aligned}
\dot{J} &= -\frac{F}{2}\left(q_1(J)e^{i\vartheta} + q_1^*(J)e^{-i\vartheta}\right) \\
\dot{\vartheta} &= \frac{\partial H_0(J)}{\partial J} - \frac{iF}{2}\left(\frac{\partial q_1(J)}{\partial J}e^{i\vartheta} - \frac{\partial q_1^*(J)}{\partial J}e^{-i\vartheta}\right) - \Omega
\end{aligned}. \tag{7}$$

Here $J(t) = \langle I(t) \rangle$ is the average of the action variable over the fast phases. It is easy to see by direct differentiation that System (7) possesses the following first integral:

$$C = H_0(J) - \frac{iF}{2}\left(q_1(J)e^{i\vartheta} - q_1^*(J)e^{-i\vartheta}\right) - \Omega J = \text{const}. \tag{8}$$

Expression (8) defines a family of 1:1 RMs. The constant $C$ is determined by initial conditions on the RM– the values of the action and slow phase, at which the system is captured by the RM. The first integral (8) is a particular case of general conservation law for the RMs of the single-DOF systems with periodically time-dependent Hamiltonian [22].

For analysis of the escape of the forced particle from the potential well, we adopt a well-known form of the potential energy [1, 20]:

$$V(x) = -\frac{V_0}{\cosh^2 \alpha x}. \tag{9}$$

This potential well is characterized by depth $V_0$, characteristic width $\alpha^{-1}$ and exponential decay of the attractive force at $x \to \pm\infty$; $x$ is a physical displacement of the particle. In terms of physical variables, the motion of particle under harmonic forcing in the well with potential (9) is described by the following equation:

$$m\frac{d^2 x}{dt^2} + \frac{2V_0 \alpha \sinh \alpha x}{\cosh^3 \alpha x} = A \sin \omega t. \tag{10}$$

Here $m$ is the mass of the particle, $A$ and $\omega$ are physical forcing amplitude and frequency respectively. Transition to the non-dimensional variables is performed as:

$$\tau = \omega_0 t,\ q = \alpha x,\ \omega_0 = \alpha\sqrt{\frac{2V_0}{m}},\ \Omega = \frac{\omega}{\omega_0},\ F = \frac{A}{2\alpha V_0}$$

Then, one obtains the equation in the form (1):

$$\ddot{q} + \frac{\sinh q}{\cosh^3 q} = F \sin \Omega \tau. \tag{11}$$

The dot denotes differentiation with respect to $\tau$. The rescaled potential energy is written as $V(q) = -\frac{1}{2}\cosh^{-2} q$. As it was mentioned above, here we deal with the model potential that belongs to the narrow class of strongly nonlinear oscillators, which can be integrated in terms of elementary functions [23]. Transition to the action-angle variables induced by such free (unforced and undamped) nonlinear oscillator is well-known [28], and is performed as follows:

$$H_0(I) = -\frac{1}{2}(1-I)^2, \quad q(I,\theta) = \operatorname{arcsinh}\left(\frac{\sqrt{2I-I^2}}{1-I}\sin\theta\right), \quad 0 \leq I \leq 1 \tag{12}$$

For the sake of completeness, the details of derivation of expressions (12) are presented in the Appendix. For the unforced system, the value $I=0$ corresponds to the bottom of the potential well, and $I=1$ - to the separatrix between the bounded and unbounded motions. Transformations (12), in accordance with general form (4), yield the following initial Hamiltonian of the problem:

$$H(I,\theta) = -\frac{1}{2}(1-I)^2 - F\sin(\Omega\tau)\operatorname{arcsinh}\left(\frac{\sqrt{2I-I^2}}{1-I}\sin\theta\right), \quad 0 \leq I \leq 1 \tag{13}$$

This expression allows immediate computation of the conservation law on the 1:1 RM for the considered problem in accordance with Equation (8):

$$C(J,\vartheta) = -\frac{1}{2}(1-J)^2 - \frac{2F}{\pi k}(\mathbf{K}(k) - \mathbf{E}(k))\cos\vartheta - \Omega J = \text{const}, \quad k = \sqrt{2J-J^2} \tag{14}$$

Here $\mathbf{K}(k)$ and $\mathbf{E}(k)$ are complete elliptic integrals of the first and the second kind, respectively, $k$ is the modulus of the elliptic integrals. Details of derivation of Expression (14) are presented in the Appendix.

3. **Structure of the RM and the escape threshold.**

The next step is to explore the qualitative structure (phase portrait) of the RM satisfying conservation law (14). It is easy to see that the fixed points of this phase portrait correspond to solutions of the following transcendental equations:

$$\frac{\partial C(J,\vartheta)}{\partial \vartheta} = 0 \Rightarrow \sin\vartheta = 0; \quad \frac{\partial C(J,\vartheta)}{\partial J} = 0 \tag{15}$$

The first equation yields that the fixed points may appear only at lines $\vartheta_e = 0, \pi$. The second equation is awkward and hardly solvable analytically; then, one should rely on a combination of numeric and analytic methods. In order to explore the escape problem, we adopt that the forcing in Equation (11) is switched at $\tau = 0$ and the initial conditions at this moment are zero: $q(0) = \dot{q}(0) = 0$. Then, one can *assume* that the particle is captured into the RM at zero, or close to zero initial conditions, and to consider the behavior of the special RM orbit, which corresponds to such initial conditions. In recent literature [17, 18, 29, 30] such orbits are referred to as limiting phase trajectories (LPTs), and this terminology will be followed. It is easy to identify the LPT, since it corresponds to the initial condition $J=0$. Then, it should satisfy the following equation:

$$C(J,\vartheta) = -1/2 \qquad (16)$$

It turns out that the qualitative structure of the RM is different in the cases of small and large forcing frequencies $\Omega$. A typical transformation of the phase portrait in the case of relatively small forcing frequencies is presented in Figure 1. All phase portraits below are plane views of the phase cylinder $\{(J,\vartheta);\ 0 \leq J \leq 1,\ 0 \leq \vartheta < 2\pi\}$.

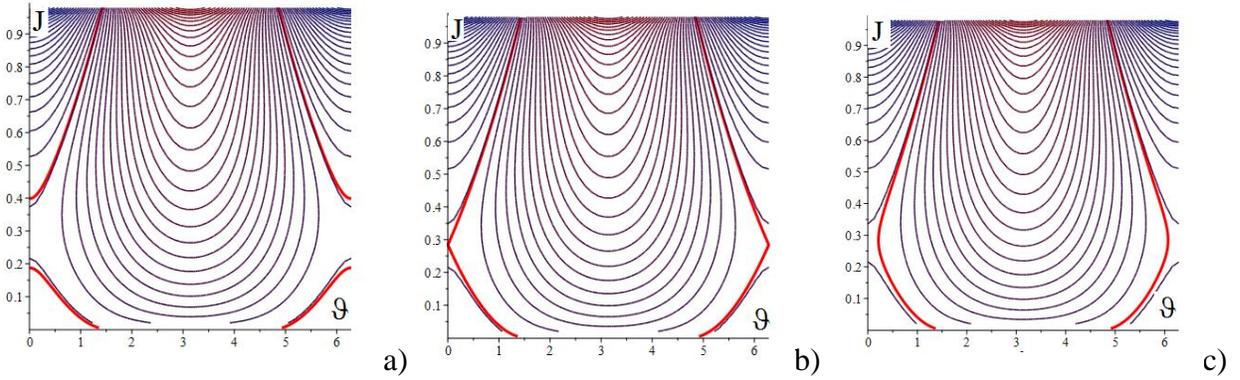

*Figure 1. Reconnection of the LPT for the case of relatively small forcing frequencies; $\Omega=0.4$, a) $F=0.28$, b) $F_{crit}=0.2934$, c) $F=0.3$. Thick red line denotes the LPT (16).*

We observe that for relatively small forcing the upper and lower branches of the LPT (16) are not connected. Thus, if the particle starts from the zero initial conditions at the RM, the value of $J$ remains relatively small, and the particle does not escape from the potential well (Figure 1a). For larger values of forcing, the LPT has only one branch, which approaches $J=1$ (Figure 1c). Physically, one should identify this situation with the escape from the potential well, since the unit value of the action corresponds to the separatrix of the unforced system. The marginal RM structure, that divides between the two scenarios, is depicted in Figure 1b. Here the LPT branches connect in the saddle point at $\vartheta_e = 0$. Therefore, the LPT coincides with the separatrix of the RM (not to be mixed with the separatrix of the unforced system). For given excitation frequency Ω, it is natural to identify the value of $F_{crit}$ that yields Figure 1b with the minimal forcing amplitude required for the escape of the particle from the potential well. We refer to this special transition as the LPT reconnection.

For the case of relatively large excitation amplitudes, one observes the other scenario of the LPT reconnection, as illustrated in Figure 2.

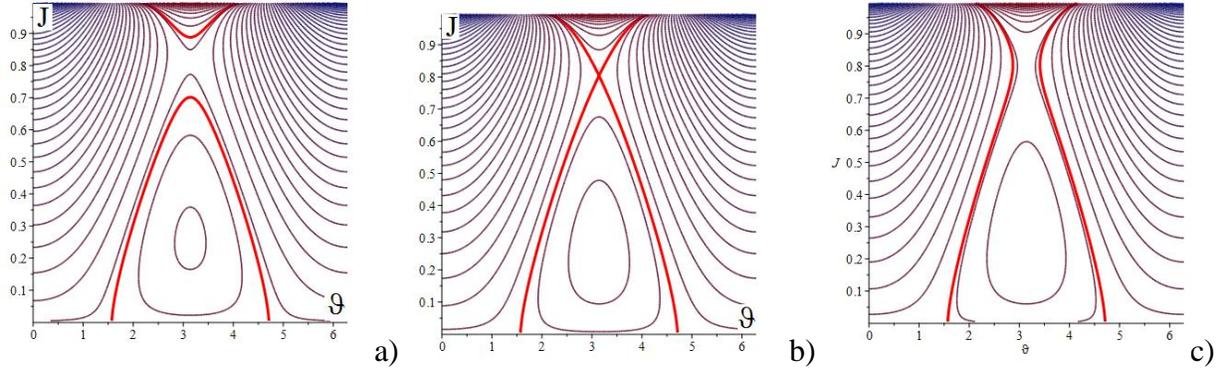

*Figure 2. Reconnection of the LPT for the case of relatively large excitation frequencies; Ω=1, a) F=0.24, b) $F_{crit}$=0.25057, c) F=0.26. Thick red line denotes the LPT (16).*

In this case, the reconnection occurs in the saddle point at $\vartheta_e = \pi$. Thus, one can identify two possible scenarios of the LPT reconnection. Each point of the LPT reconnection determines the pair $(F_{crit}, \Omega)$, where $F_{crit}$ is the minimal theoretical forcing amplitude required for the escape. Two scenarios of the LPT reconnection lead to two curves $F_{crit}(\Omega)$ in the parametric plane $(F,\Omega)$.

It seems impossible to derive the explicit analytic dependence $F_{crit}(\Omega)$, but the implicit parametric expression is relatively easy to obtain. To perform that, it is enough to recall that the reconnection occurs at the saddle points. Therefore, the values of $F_{crit}$ for given values of Ω should satisfy the following equations:

$$\frac{\partial C}{\partial J} = 0 \Rightarrow (1-J) \pm F_{crit} G_1(J) - \Omega = 0$$

$$C(J,\vartheta) = -\frac{1}{2} \Rightarrow J - \frac{J^2}{2} \pm F_{crit} G_0(J) - \Omega J = 0 \qquad (17)$$

$$G_0(J) = \frac{2}{\pi k}(\mathbf{K}(k) - \mathbf{E}(k)), \; G_1(J) = \frac{\partial G_0(J)}{\partial J} = \frac{2(1-J)}{\pi}\left(\frac{\mathbf{E}(k) - (1-k^2)\mathbf{K}(k)}{k^3(1-k^2)}\right), \; k = \sqrt{2J - J^2}$$

The first equation of System (17) follows from the requirement to pass through the fixed point; the second one defines the LPT. Positive and negative signs correspond to $\vartheta_e = \pi$ and $\vartheta_e = 0$ respectively. The fixed point in question must be the saddle, since it is the only generic possibility for the phase trajectory to pass through it.

Equations (17) can be easily solved, and yield the following parametric expressions for $F_{crit}(J)$ and $\Omega(J)$:

$$F_{crit}(J) = \mp \frac{J^2}{2(G_0(J) - JG_1(G))}; \; \Omega(J) = \frac{G_1(J)(J^2 - 2J) - 2G_0(J)(J-1)}{2(G_0(J) - JG_1(G))}. \qquad (18)$$

The curves defined by Equations (18) in physically meaningful part of the parametric plane $F_{crit} > 0, \Omega > 0$ are presented in Figure 3.

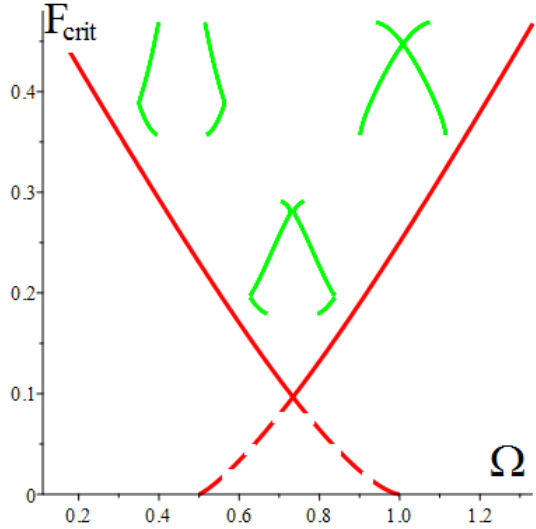

Figure 3. The LPT reconnection points in the (F,Ω) plane. The ascending and descending curves correspond to the positive and negative signs in (18) respectively. The scheme near each curve

*describes the LPT reconnection scenario that corresponds to this curve. The lower scheme corresponds to the intersection point.*

One can observe that every scenario of the LPT reconnection exists in a certain range of the excitation frequencies. The parts of the curves in Figure 3 below the intersection point still correspond to the LPT reconnections, but do not lead to the escape – the second saddle point prevents it. The intersection point of the curves in Figure 3 is of substantial physical interest – it corresponds to minimal theoretical amplitude of the forcing, at which the escape from the well is still possible. Numeric solution yields the following coordinates for this point:

$$\Omega^* = 0.73395, \quad F^*_{crit} = 0.09659 \tag{19}$$

For these special values of the frequency and the forcing amplitude, the RM phase portrait has a peculiar shape presented in Figure 4.

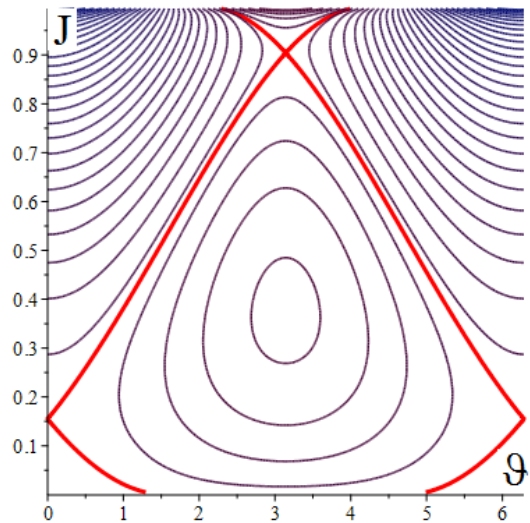

F*igure 4. Double reconnection of the LPT for special values of parameters (19).*

One observes that the LPT coincides with a heteroclinic orbit that connects two saddle points.

The theoretical exploration presented above predicts the non-monotonous dependence of the minimal forcing amplitude required for the escape, on the excitation frequency – the graph in

Figure 3 exhibits a minimum associated with the double LPT reconnection. In order to validate these predictions, we perform direct numeric simulations of System (11)

## 4. Numeric validations and illustrations.

First, we present a typical escape scenario for values of parameters defined in the figure caption (Figure 5).

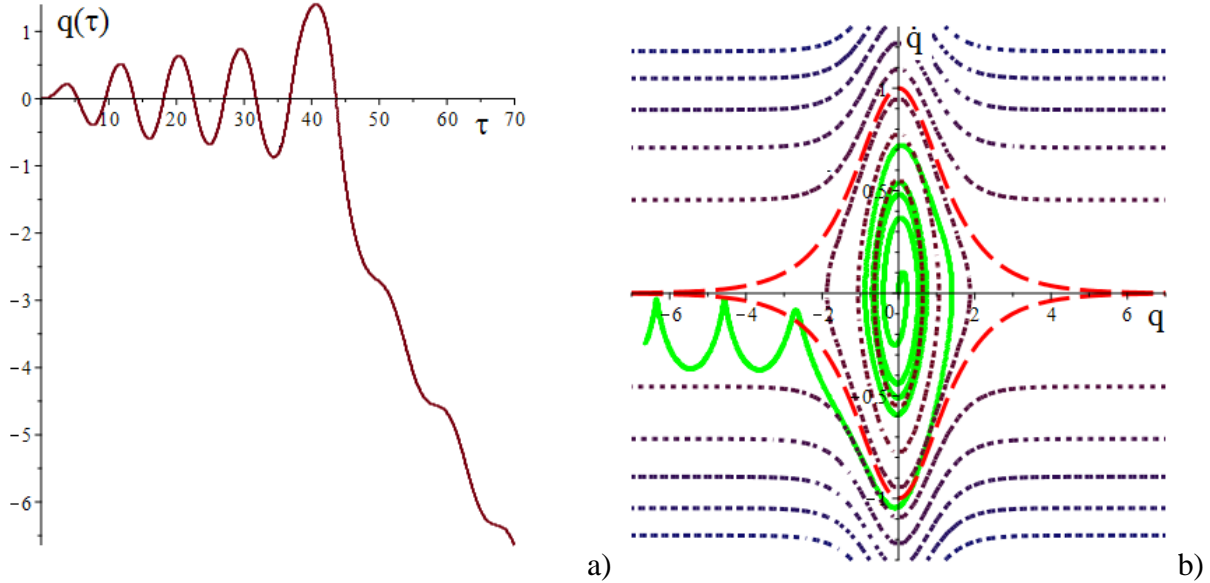

*Figure 5. Time series for solution of equation (11) with zero initial conditions and $\Omega = 0.7, F = 0.119$; a) time series $q(\tau)$; b) solution in $(q, \dot{q})$ plane superimposed on the phase portrait of the unforced system. Separatrix of the unforced system is denoted by dashed line, other phase trajectories – by dotted lines. Solid thick line corresponds to solution of the forced system $(q(\tau), \dot{q}(\tau))$.*

The escape process presented in Figure 5 starts from clear transient resonance pattern ($0 < \tau < 45$). Then, the trajectory follows the separatrix (cf. Figure 5b), and finally escapes the well at $\tau \approx 50$. It is possible to suggest that the separatrix crossing also can affect the escape process; this effect lies beyond the RM formalism presented above.

Comparison of numerically determined minimal values of the forcing necessary for the escape, to the theoretical predictions is presented in Figure 6.

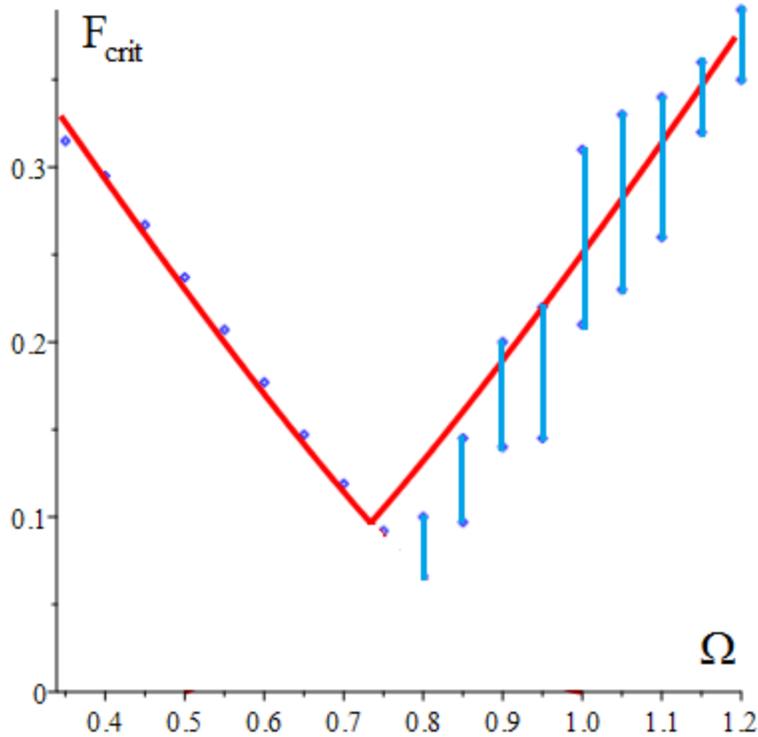

*Figure 6. Comparison of numerically determined minimal values of forcing, necessary for the escape, to the forcing frequency. Solid line corresponds to the theoretical predictions (cf. Figure 3), fragments of the curves below the intersection points are removed. Diamonds denote the minimum forcing values determined numerically. In the cases where the interval of the forcing values is presented, the upper diamonds denote the forcing that provides escape at $\tau \leq 200$. The lower diamonds correspond to the escape at $\tau \leq 2000$.*

The most significant observation from Figure 6 is the existence of the minimum at $\Omega - F_{crit}$ curve, in agreement with the theoretical predictions. Moreover, this minimum is reasonably close to point (19). General shape of the curve is also similar to the one predicted by the theory.

In the same time, the numeric simulation reveals some important features that are overlooked by the presented theoretical approach. First of all, the minimal forcing level is somewhat vaguely defined. For the range of $1 > \Omega > \Omega^*$ the numeric value of $F_{crit}$ strongly depends on the time span of the simulation. In Figure 6, somewhat voluntarily, the lower point corresponds to the escape observed for relatively long time $\tau \leq 2000$ and the upper one − for relatively rapid escape on the time span $\tau \leq 200$. Different definitions of the time spans would yield somewhat different intervals of the $F_{crit}$. Moreover, for some values of the parameters between these boundaries, it can happen that the escape is not observed even for $\tau \geq 2000$. These peculiarities signify the chaotic-like behavior of the particle, related to the separatrix crossing. The latter is inevitable for the escape, but, in the same time, it could introduce essential chaotic features in the dynamics [31, 32]. One can speculate that this effect is more profound for the large frequencies, since in this case the theoretical minimal forcing is governed by the LPT reconnection at the upper saddle. This saddle is relatively close to the separatrix, and therefore the effect of the latter on the transition is more profound in this case. Then, the ascending part of the theoretical curve in Figure 6 should be treated as a general trend, rather than as the rigorous prediction. To the contrary, for $\Omega < \Omega^*$ the numeric transition to the escape is rather sharp, and the coincidence between the theoretical predictions and the numeric results is almost perfect.

In order to reveal and verify directly the resonance capture scenario, we simulate the instantaneous frequency of the response as a function of time, and compare it to the excitation frequency. For this sake, we solve numerically the following exact dynamical equations derived from Hamiltonian (13), with appropriate initial conditions:

$$\dot{\theta} = \frac{\partial H(I,\theta)}{\partial I} = 1 - I - \frac{F\sin(\Omega t)\sin(\theta)}{\sqrt{2I-I^2}(1-I)^2\sqrt{1+\frac{2I-I^2}{(1-I)^2}\sin^2\theta}},$$

$$\dot{I} = -\frac{\partial H(I,\theta)}{\partial \theta} = \frac{\sqrt{2I-I^2}F\sin(\Omega t)\cos(\theta)}{(1-I)\sqrt{1+\frac{2I-I^2}{(1-I)^2}\sin^2\theta}} \quad . \quad (20)$$

$$0 < I < 1$$

The first equation of System (20) delivers an expression for the instantaneous frequency of the particle oscillations. Typical result of the simulations is presented in Figure 7.

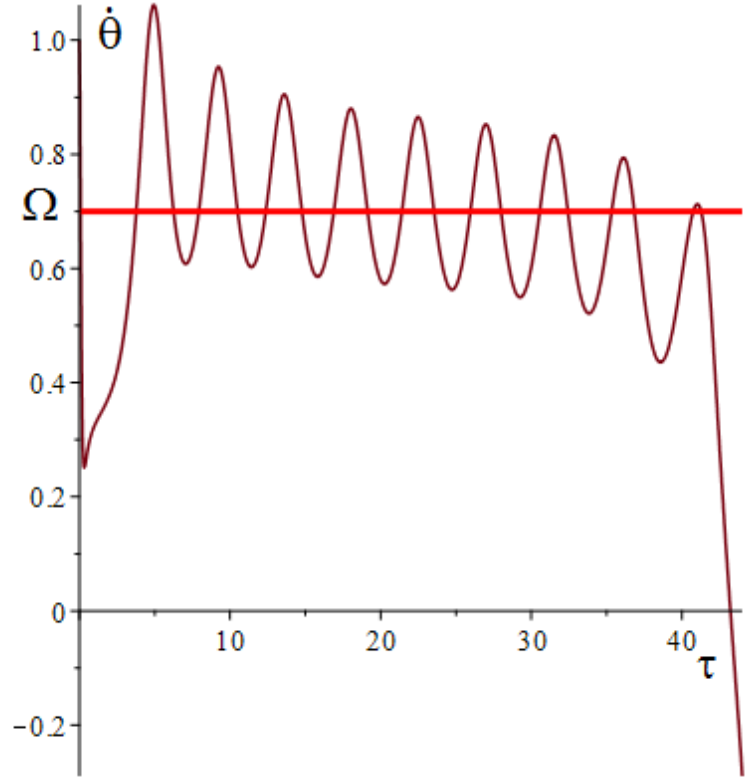

*Figure 7. Instantaneous frequency of the particle, evaluated from the first equation of System (20), versus time. Thick line corresponds to the excitation frequency, $\Omega = 0.7, F = 0.119$.*

The result presented in Figure 7 confirms the scenario of the capture into the RM – one observes that the particle instantaneous frequency oscillates around the excitation frequency up to the escape point.

### 5. Concluding remarks.

Main result of the work, presented in Figure 6, namely, the sharp minimum of the critical external forcing for $\Omega = \Omega^*$, conforms to numeric and semi-analytic results obtained in [4] and other related works for broad variety of the well potentials. The values of the critical frequency, reported there, vary approximately in the interval 0.75-0.85 of the frequency of free oscillations at the bottom of the well (the exact value depends on the model example and on the empirical safety factor used). Then, it is possible to conjecture that this pattern is universal for generic escape problem under the periodic forcing. Moreover, it is reasonable to suggest that the sharp minimum in other models also appears due to the competition between two topological mechanisms of the LPT reconnection. Still, this conjecture requires additional verification and proof. The results also point on advantage of the energy-based methods for estimation of the escape. Free oscillator with potential energy (9) does not possess the saddle point, so the notion of critical displacement seems meaningless, or at least vaguely defined. Still, qualitative response of this model is similar to other models, in which such saddle points exist.

The above conjectures require clarifying the role of the damping, since the models used in [4] and in other related papers included the damping terms. The damping will not allow derivation of the conservation law similar to (14). This conservation law substantially simplifies the exploration of the RM structure. However, the lack of this integral, and, in general, the lack of the Hamiltonian structure does not prevent the transformation to the AA variables and subsequent exploration of the dynamics [25, 26]. The averaging can be performed directly in the equations of motion (cf. (20)), and, again, the average equations will describe the motion on cylindrical RM $(J, \vartheta)$. Special phase trajectories corresponding to various initial conditions on the RM also are easy to identify. The results presented above demonstrate that the transitions to the escape are governed by the LPT reconnection scenarios. The latter, in turn, are governed by the saddle points on the RM. These saddle points, due to their hyperbolic character, will be only slightly perturbed, if the RM will be slightly modified because of the small damping. Therefore, one can expect that the generic LPT reconnection scenarios, leading to the escape, will reveal themselves also in the damped systems with generic damping functions, if the damping will be weak enough. From the other side, a global effect of the damping can be quite significant, especially if the escape time will be large. In particular, one can expect that the damping will suppress, at least to some extent, the chaotic phenomena related to the separatrix crossing. All these issues require additional exploration.

### Acknowledgment.

The author is very grateful to Israel Science Foundation (grant 838/13) for financial support.

### Appendix.

This appendix presents the derivation details for expressions (12) and (14). In accordance with definition (3), the action variable is expressed as follows:

$$I(E) = \frac{1}{2\pi}\oint p(q,E)dq = \frac{1}{2\pi}\oint\sqrt{2E+\frac{1}{\cosh^2 q}}dq = \frac{1}{2\pi}\oint\sqrt{2E+1-\tanh^2 q}\,dq =$$

$$= \bigg|_{a=\sqrt{1+2E},\tanh q=z} = \frac{2}{\pi}\int_0^a \frac{\sqrt{a^2-z^2}}{1-z^2}dz = 1-\sqrt{1-a^2} = 1-\sqrt{-2E} \qquad (A1)$$

The last integral is easily evaluated with standard methods of contour integration. Inversion of (A1) yields the expression for $H_0(I)$ from (12). Expression for the coordinate is calculated as follows:

$$\theta = \frac{\partial}{\partial I}\int_0^q p(q,I)dq = \frac{\partial}{\partial I}\int_0^q\sqrt{2E+\frac{1}{\cosh^2 q}}dq = \frac{\partial}{\partial I}\int_0^q\sqrt{\frac{1}{\cosh^2 q}-(1-I)^2}\,dq =$$

$$= -(1-I)\int_0^q \frac{dq}{\sqrt{\frac{1}{\cosh^2 q}-(1-I)^2}} = -(1-I)\int_0^q \frac{d(\sinh q)}{\sqrt{2I-I^2-(1-I)^2\sinh^2 q}} = \qquad (A2)$$

$$= -\arcsin\left(\frac{(1-I)\sinh q}{\sqrt{2I-I^2}}\right)$$

Then, we obtain the second expression of (12), with insignificant change of sign. By symmetry considerations, Expression (12) for $q(I,\theta)$ can be expanded in sine-Fourier series:

$$q(I,\theta) = \operatorname{arcsinh}\left(\frac{\sqrt{2I-I^2}}{1-I}\sin\theta\right) = \sum_{m=1}^{\infty} a_m \sin m\theta \qquad (A3)$$

From (8) it is obvious that it is necessary to compute only the coefficient $a_1$. One obtains:

$$a_1 = \frac{1}{\pi}\int_{-\pi}^{\pi}\operatorname{arcsinh}(c\sin\theta)\sin\theta\,d\theta = -\frac{1}{\pi}\operatorname{arcsinh}(c\sin\theta)\cos\theta\bigg|_{-\pi}^{\pi} + \frac{4c}{\pi}\int_0^{\pi/2}\frac{\cos^2\theta}{\sqrt{1+c^2\sin^2\theta}}d\theta =$$

$$= \bigg|_{\cos\theta=\xi,k=\frac{c}{\sqrt{1+c^2}}} = \frac{4}{\pi}\int_0^1 \frac{\xi^2 d\xi}{\sqrt{1-\xi^2}\sqrt{1/k^2-\xi^2}} = \frac{4}{\pi k}(\mathbf{K}(k)-\mathbf{E}(k)) \qquad (A4)$$

Here $c = \frac{\sqrt{2I-I^2}}{1-I}$, and, consequently, $k = \sqrt{2I-I^2}$, as stated above. Substituting (A3-A4) into (5) and (8), one obtains $q_1 = -\frac{ia_1}{2}$ and then - expression (13). Further coefficients

$a_m, m > 1$ also are combinations of rational functions and complete elliptic integrals, but become quite awkward even for small *m*.